\documentclass[a4paper,fleqn,usenatbib]{mnras}

\usepackage[T1]{fontenc}
\usepackage{ae,aecompl}

\usepackage{graphicx}	
\usepackage{amsmath}	
\usepackage{amssymb}	

\title[A MAD model of GRB variability]{A MAD Model for Gamma-Ray Burst Variability}

\author[N.M. Lloyd-Ronning et al.]{
Nicole M. Lloyd-Ronning,$^{1,2}$\thanks{E-mail: lloyd-ronning@lanl.gov}
Joshua C. Dolence,$^{1,2}$
and Christopher L. Fryer$^{1,2}$
\\
$^{1}$ CCS-2, Los Alamos National Laboratory, Los Alamos, NM, 87545, USA\\
$^{2}$Center for Theoretical Astrophysics, Los Alamos National Laboratory, Los Alamos, NM, 87545, USA\\
}

\pubyear{2016}

\begin{document}
\label{firstpage}
\pagerange{\pageref{firstpage}--\pageref{lastpage}}
\maketitle

\begin{abstract}
 We present a model for the temporal variability of long gamma-ray bursts during the prompt phase (the highly variable first 100 seconds or so), in the context of a magnetically arrested disk (MAD) around a black hole.  In this state, sufficient magnetic flux is held on to the black hole such that it stalls the accretion near the inner region of the disk.  The system transitions in and out of the MAD state, which we relate to the variable luminosity of the GRB during the prompt phase, with a characteristic timescale defined by the free fall time in the region over which the accretion is arrested.  We present simple analytic estimates of the relevant energetics and timescales, and compare them to gamma-ray burst observations.  In particular, we show how this model can reproduce the characteristic one second time scale that emerges from various analyses of the prompt emission light curve.   We also discuss how our model can accommodate the potentially physically important correlation between a burst quiescent time and the duration of its subsequent pulse (Ramirez-Ruiz \& Merloni 2001).    
\end{abstract}


\section{Introduction}
 The variability of a gamma-ray burst (GRB) was one of the first observables that provided a glimpse into the nature of their progenitors.  During the first 100 seconds or so (during the so-called prompt phase), GRBs emit most of their energy in gamma-rays, with a flux that varies on $\le 1s$ timescales.  This rapid variability gives an indication of the size of the emission region ($ \sim c \delta t \Gamma^{2}$, where $\Gamma$ is the Lorentz factor of the flow), and allows us to associate GRBs with compact stellar objects.  A huge leap in our understanding came in the late 1990s with redshift measurements to GRBs, which gave a determination of the general energetics and led to the development of viable progenitor models.  

There is reasonable evidence (see, e.g., Gehrels, Ramirez-Ruiz \& Fox 2009, Berger 2014) to support the claim that long GRBs (duration $\ge 2 s$) are associated with the collapse of a massive star, while short GRBs (duration $\le 2s$) are associated with the merger of two compact objects.  In both cases, a black hole can be created with a surrounding disk (as we discuss below, however, there are a number of models in which the central object is a  supramassive NS, with a delayed collapse to a black hole; for example, see Bucciantini 2012, Lu \& Zhang 2014, Gao et al. 2015).  This black hole-disk system launches a jet, likely through either neutrino-antineutrino annihilation (Liu et al. 2015), or a Blandford-Znajek process (Blandford \& Znajek 1977) in which the rotational energy of the black hole is extracted via a poloidal magnetic field threading the horizon of the black hole.  The Blandford-Znajek process more easily provides the necessary power in the jet needed to produce the observed GRB, and is regarded as the likely jet launching mechanism (Lee et al. 2000, Liu et al. 2015, Kawanaka, Piran and Krolik 2013, Tchekhovskoy et al. 2011, McKinney et al. 2012).

\subsection{The Nature of Gamma-Ray Burst Variability}
   A long-standing problem in the field of gamma-ray bursts has been uncovering the nature of the observed variability during the prompt gamma-ray phase.  Various authors have addressed this problem with different conclusions.  One viable way to produce the initial variable burst of gamma-rays, and its long-lived afterglow is via a radially modulated outflow.  For example, in the internal-external shock paradigm (see, e.g., Piran 2004 for a review), the jet launches multiple shells of matter with different Lorentz factors.  Faster shells catch up with slower shells and create so-called internal shocks. Multiple shocks lead to multiple emission episodes, which can account for the rapidly fluctuating light curve of a prompt GRB (however, shocks are not necessarily the dominant source of dissipation in GRBs; what we discuss below is generic to any observable radially modulated outflow, and not specific to a particular dissipation mechanism).

 Kobayashi et al. (1997) and Nakar \& Piran (2002) showed that for efficient collisions of shells with random bulk Lorentz factors, the observed peaks are in nearly one-to-one correspondence with the activity of the central engine.  This can be understood as follows:  Consider two shells with a separation $L$ emitted at time $t_{1}$ and $t_{2} \sim t_{1} + L/c$.  The shells will collide at a radius $R_{coll} \approx 2 \Gamma^{2}L$.  The photons from this collision will reach the observer at a time $t_{obs} \approx t_{1} + R_{coll}/2\Gamma^{2} \approx  t_{1} + L/c \approx t_{2}$.  In other words, the observed time of the collision reflects the emission of the (faster) shell from the central engine.   Therefore, in the context of the simple picture of multiple shell ejection in the jet, one reasonable interpretation is that the observed GRB variability is {\it a direct reflection of the central engine variability}. 

 Others have argued for a different origin for the GRB variability.  Thompson \& Gill (2014) consider a hot, magnetized jet that interacts with a confining medium.  The GRB emission arises when the magnetofluid is able to break out of the confining baryon shell and expand relativistically, which occurs when a corrugation instability develops after a period of time (greater than the duration of the central engine).  Each GRB pulse corresponds roughly to the total active period of the central engine ($\sim 1s$ in their model), while the duration of the observed burst is a reflection of the angular timescale over which the regions of magnetic breakout are observed.  Barniol Duran et al. (2015)  explain the prompt GRB light curve through so-called mini-jets - small scale jets that move relativistically with respect to the overall outflow.  In their model, the minijets - possibly physically interpreted as reconnection regions - are oriented perpendicular to the outflow, and dissipate at a radius $R_{diss} = \Gamma^{2} c t_{GRB}$, where $t_{GRB}$ is the observed duration of the burst.  Their model can achieve not only a high degree of variability, but allows for a steep decline at the end of the prompt emission (in accordance with the observations).  
   
 However, even in the case of radially modulated outflow from the central engine, it is not obvious that the central engine variability is reflected in the observed light curve. Ramirez-Ruiz et al. (2001) have shown that variability can in fact arise in certain cases where the central engine is {\em not} switching on and off. They look specifically at quiescent times (long intervals between pulses - see \S 2 below) in GRBs, and find that these pauses in a prompt GRB light curve are not necessarily indicative of central engine dormancy.  There are two cases in which this is possible: 1) if the shells in the jet are ejected all at the same Lorentz factor, or 2) if the shells decrease in Lorentz factor with each ejection episode.  In either case, subsequent shells (emitted later) will not catch up with previously ejected shells until well into the deceleration phase of the blast wave, and the observed quiescent times between pulses can be reproduced.  

  Besides the fact that these two scenarios require perhaps somewhat contrived conditions for the shell ejection, we would expect to see evidence of these models in the GRB light curve. For example,  evidence of deceleration should appear in the prompt GRB light curve in the form of a systematic increase of pulse width as a function of time, which is not seen in most cases (other observational signatures of a wind in these two cases are discussed in  Ramirez-Ruiz et al. 2001).  Drago \& Pagliara (2007) performed a statistical analysis of the temporal structure of long gamma-ray bursts, and found that the average duration of a quiescent time is comparable to the average duration of an emission episode.   In addition, they found pre- and post-quiescent emission exhibit similar properties such as temporal structure, spectral hardness ratios and emitted powers.  They interpret their results as an indication that quiescent times reflect periods of dormancy in the inner engine.  It therefore seems reasonable to construct a model from the standpoint that, for the case of a radially modulated outflow, the observed variability is a direct reflection of the the central engine variability.  {\em However, the physical nature of this variability at the central engine remains unknown. }  

  This is the question we attempt to address in this paper. We present a model for GRB variability (for long GRBs lasting $> 2s$) originating at the progenitor. We invoke a magnetically arrested disk (MAD) in which variability arises naturally from unstable, variable accretion flow mediated by interchange instabilities and/or reconnection events. The variability timescale is directly related to the free fall time in the magnetically arrested portion of the disk. The format of the paper is as follows: In \S 2, we give an overview of relevant observed GRB timing properties.  In \S 3, we describe our model and give analytic estimates of the timescales and energetics, and compare these to observed GRB properties.  In particular, in \S 3.2, we show that for reasonable disk parameters, the free fall time in a magnetically arrested disk is on the order of the characteristic GRB variability time $\sim 1s$. We discuss some further implications of our model in terms of late time activity of GRBs, as well as some caveats of our model in \S 4.  Conclusions are presented in \S 5.

\section{Observed GRB variability/timing properties}
  There have been many detailed investigations of  specific timing properties of GRBs (again, here, we discuss only the results referring to the majority of bursts - long GRBs with durations $> 2s$), with the goal of uncovering the underlying physical mechanism (e.g., Norris et al. 1996, Quilligan et al. 2002, McBreen et al. 2002, and additional references below).  Significantly, a characteristic timescale of order $\sim 1$ second seems to emerge from many of these studies, although many different time analysis techniques were employed.  Below, we give a brief summary of the results from these studies, highlighting the characteristic time scales and physically relevant correlations that have emerged.

\begin{enumerate}
\item{\underline{Pulse width distribution}:}
 Several authors (Norris et al. 1996, Quilligan et al. 2002, Nakar \& Piran 2002) have analyzed the durations or widths of individual pulses in GRBs.  Although an individual pulse is not always well-defined, these authors have employed various pulse-fitting algorithms with similar conclusions.  The distribution of pulse widths in GRBs is a log-normal, extending from 0.1 to 10 seconds, and peaking at around 1 second. We point out that the minimum timescale for a GRB pulse is limited by the angular time scale (the time over which photons from a shell at radius $R$, will reach the observer), $T_{ang} = R/2\Gamma^{2}$, where $\Gamma$ is the Lorentz factor of the outflow.   This can produce a turnover in the pulse width distribution that may not necessarily reflect a physical timescale at the central engine.

\item{\underline{Variability timescale}:}
   The variability of a GRB is characterized by some measure of the average deviation of the light curve relative to a smoothed time profile, and can give an indication of how multi-peaked a GRB light curve is on some timescale. GRBs can exhibit variability on many different timescales (e.g. there is often a rapid noise-like variability on top of a smoother modulation of the time profile), and a number of groups have employed various methods to rigorously define variability (Fenimore \& Ramirez-Ruiz 2000, Reichart et al. 2001,  Guidorzi 2005). Margutti (2009) analyzed the maximum-variability timescale -  the time scale over which the signal shows the maximum variability power - in a number of GRBs, both in the observer frame and in the cosmological rest frame of the burst.  Using a sample of 75  bursts, they found a maximum variability timescale {\em in the source frame} varying from 0.1s to a few seconds, peaking strongly at about 1 second.  This peak at $1s$ is identified as a characteristic timescale in the source frame.
  
\item{\underline{Power spectral density}:}
  Another way to analyze the GRB time profile is to compute its power spectral density (PSD).  Beloborodov et al. (2000) showed that the average PSD for a large number of BATSE bursts is a power-law with index  $-5/3$, and a sharp break $\sim 1$ Hz.  This was confirmed by Dichiara et al. (2013) who analyzed 205 GRBs detected by the Fermi satellite and 67 GRBs detected by BeppoSAX and also found a significant break in the PSD at $\sim 1$ Hz. As pointed out in Beloborodov et al., if this timescale were produced in the rest frame of the flow (as opposed to at the central engine), it requires an unlikely distribution of Lorentz factors (very narrow with $\Delta \Gamma \sim 2$) of the multiple shells of ejecta.  It is also possible this break at 1 second is associated with the inner radius of the optically thin region of the GRB, $t \sim R/c \Gamma^{2}$, but  that leads to the difficulty of explaining why all bursts have the same characteristic $t$.  The simplest explanation for this break is that {\it it reflects a characteristic timescale of the central engine}.

\item{\underline{Distribution of intervals between pulses}:}
   Several groups (McBreen et al. 1994, Li \& Fenimore 1996, Norris et al. 1996, Quilligan et al. 2002, Nakar \& Piran 2002, Ramirez-Ruiz, et al. 2002) have analyzed the distribution of time between pulses in the prompt GRB light curve.  As with variability, there are many ways to quantify this timescale and there is always the issue that an apparent interval time is in fact just a weak signal that has not risen above the noise (and therefore not a true lack of signal).  However, using various algorithms to define and analyze the time between GRB pulses, these studies have arrived at similar conclusions. If long, so-called quiescent times (typically 10's of seconds time interval between pulses) are eliminated, the distribution of times between pulses is a log-normal peaking at around 1 second (see, for example, Figure 6 of Nakar \& Piran 2002).  
  
\item{\underline{Quiescent times}:}
Quiescent times - roughly defined as long periods of time (usually $\geq 10s$) between pulses - have also been analyzed by various groups (Ramirez-Ruiz et al. 2001, Nakar \& Piran 2002, Drago \& Pagliara 2007).   As mentioned in the introduction, these quiescent times can arise either from central engine dormancy or from a carefully modulated wind.  Each of these studies employed various techniques to arrive at the conclusion that a GRB quiescent time is more likely a result of central engine dormancy.
  A potentially very relevant clue to the nature of the central engine variability is the linear relation between the quiescent time {\bf before} a pulse and the duration of that pulse, first found by Ramirez-Ruiz \& Merloni (2001) (see also  Nakar \& Piran 2002, Drago \& Pagliara 2005).  When connecting to the central engine, such a correlation suggests the longer energy is stored at the central engine, the longer it will dissipate (as opposed to a system in which the strength of a dissipation episode determines the {\it subsequent} length of time it takes to re-accumulate enough energy for the next dissipation episode; in this case there would be a correlation between the duration of a pulse and the quiescent time {\bf after} that pulse).  This may be a powerful constraint when constructing any model of inner engine variability.   
\end{enumerate}

\section{The Model - Magnetically Arrested Disk}
  Under the assumption that GRB variability reflects central engine variability, we present a model of the central engine based on a magnetically arrested accretion disk (MAD). The basic idea behind a MAD is straightforward:  a significant amount of poloidal field (either generated via a dynamo mechanism - see, e.g., M{\"o}sta et al. 2015, Akiyama et al. 2003 - or present from the conservation of magnetic flux during the collapse of the progenitor`s extant field) is dragged in by the gas as it accretes on to the black hole.  The flux that accumulates at the horizon provides significant pressure against the infalling gas, which arrests the accretion flow within a certain radius $R_{m}$ outside the event horizon.  In this region, the velocity of the gas is less than the free-fall velocity and accretion is highly variable, occurring when the flow undergoes an interchange instability or reconnection events (Narayan et al. 2003, Spruit \& Taam 1990).  There is evidence for dynamically important magnetic fields leading to a MAD state in supermassive black holes in radio loud active galaxies (Zamaninasab et al. 2014).  Here, we consider a MAD state in the context of gamma-ray bursts.  Note that Tchekhovskoy \& Giannios (2015) have also invoked MAD models in the context of GRBs - in their model, the MAD state corresponds to the abrupt shut-off of the central engine, at a point when the continuously falling accretion rate drops low enough to where the flux on the hole becomes dynamically important (i.e. the MAD state; see their Figure 7).   In our model, the disk-black hole is in the MAD state {\it during}  the GRB, and is the {\em reason for the observed GRB variability}.

  In our simple picture, a Poynting dominated jet is launched via a Blandford-Znajek process, while the magnetic flux is anchored to the black hole.  When enough matter builds up in the disk to overcome the pressure of the field, an interchange instability can occur (Spruit et al. 1995, Spruit \& Taam 1990, Ikhsanov 2001).  This instability is analogous to a Rayleigh-Taylor instability - the density of the accreting matter reaches a point in which it is energetically favorable to cross the field lines, while the magnetic flux diffuses out into the disk (see \S 3 of Spruit \& Taam, 1990, where microphysical, not turbulent, diffusion is considered).  The growth rate of the interchange instability is on the order of the free fall time if the ratio of surface density to poloidal field changes significantly enough as a function of radius (see Spruit \& Taam 1990 for details).   Alternatively, as the field is distorted from the pressure of the gas, reconnection events can occur and again the magnetic flux diffuses out.  The back-and-forth between the magnetic flux held to the hole and diffusing out into the disk corresponds roughly to the GRB pulse and quiescent times, respectively.
   
  It is possible that when the interchange instability occurs, some of the mass crossing the field lines is ejected out into the jet as a shell of matter (as in the internal-external shocks scenario), and subsequently accelerated by the Poynting jet.  Whether it is the collision of these ``shells``, or the modulated Poynting flux (discussed above) that produces the resultant GRB emission, our estimates for the energetics and timescales involved in this problem remain the same.   We discuss the differences in these two possible scenarios in the context of the Ramirez-Ruiz \& Merloni correlation in \S 3.3.

\subsection{Energetics} 
 We can make some simple estimates of the field anchored to the black hole, the Blandford-Znajek luminosity and the relevant timescales in our toy model.  For the purposes of these estimates, following Narayan et al. (2003), we consider a simple picture of a non-rotating disk of gas falling into the black hole.  We can estimate the strength of the magnetic field needed to support the infalling gas:

\begin{equation} 
 \mathrm{G}M\Sigma/R^{2} \sim B_{z}^{2}/2\pi
\end{equation}

\medskip

\noindent where $\Sigma = \dot{M}/(2\pi R \epsilon v_{ff})$ is the surface density of the gas, $\mathrm{G}$ is the gravitational constant, $\epsilon v_{ff}$ is the arrested velocity of the flow (compared to the free-fall velocity $v_{ff} = \sqrt{2\mathrm{G}M/R}$), and $\dot{M}$ is the accretion rate. The value of $\dot{M}$ is an unknown quantity, but various studies of gamma-ray transients from progenitor disk systems (e.g. Woosley \& Heger 2012; Lindner et al. 2010) suggest that $\dot{M}$ can range from as high as $0.1 M_{\odot} s^{-1}$ down to $< 10^{-4}M_{\odot} s^{-1}$ below which the accretion is presumably too low to power the GRB during the prompt phase.  We normalize the accretion rate in our expressions below to the fairly conservative value of $10^{-2} M_{\odot} s^{-1}$.  The parameter $\epsilon$ is on less firm footing.  As mentioned above, $\epsilon$ is a measure of how ``arrested`` the accretion flow is during the MAD state, and is a way to parameterize our ignorance of the details of the disk microphysics. It is the ratio of the radial velocity (in the MAD portion of the disk) to the free fall velocity.  A number of studies that have investigated details of diffusion through magnetic interchanges and reconnection have found analytic estimates of  $\epsilon$ that fall in the range$ \sim 10^{-2} -  10^{-3}$ (and possibly even smaller for more realistic, rotating flows; Narayan et al. 2003, Elsner \& Lamb 1984, Kaisig, Tajima, \& Lovelace 1992, Ikhsanov 2001).   Simulations of MAD disks (e.g. Igumenshchev et al. 2003, McKinney et al. 2012) find a higher value for $\epsilon$ closer to $\sim 10^{-1}$.   It may be that the discrepancy between the analytic estimates and the simulations arise from the numerical resistivity in the simulations, which causes an overestimate of the magnetic diffusion and therefore a higher $\epsilon$.  

 In what follows, we choose a number that falls in the middle of the ranges of estimates; we normalize $\epsilon$ to $10^{-2}$, keeping in mind that it can be considered a free parameter.  In this case the strength of the poloidal field can be written:

\medskip

\begin{equation}
B_{z}  \simeq 10^{16}G (\frac{5M_{\odot}}{M}) (\frac{\dot{M}}{10^{-2}M_{\odot}/s})^{1/2} (\frac{10^{-2}}{\epsilon})^{1/2} (\frac{R_{g}}{R_{m}})^{5/4}
\end{equation}

\medskip

\noindent where $R_{g}$ is the gravitational radius and $R_{m}$ is the radius to which the arrested accretion flow extends.  The presence of magnetic flux held to the black hole launches a jet via the Blandford-Znajek (BZ) mechanism (Blandford \& Znajek 1977).  The luminosity of this jet is given by the expression: 

\medskip
\begin{equation}
L_{BZ} = \frac{k f}{4\pi c} \phi^{2} \Omega^{2}
\end{equation}
\medskip

\noindent where $k$ reflects the field geometry and $\approx 0.05$ (Blandford \& Znajek, 1977;  Tchekhovskoy et al. 2010), $f$ is of order unity, $c$ is the speed of light,  $\Omega = ac/2R_{g}$ is the angular frequency at $R_{g}$, $a$ is the black hole spin parameter, and $\phi$ is the magnetic flux at $R_{g}$, $\phi = \int \int B \cdot da \sim B_{z} R_{g}^{2}$.  For a maximally rotating black hole, and using the magnetic field from equation 2 to estimate the flux $\phi$, we find

\medskip
\begin{equation}
L_{BZ} \simeq 10^{52}erg/s (\frac{\dot{M}}{10^{-2}M_{\odot}/s}) (\frac{10^{-2}}{\epsilon}) 
\end{equation}
\medskip

\noindent GRB pulses are observed to have luminosities $\sim 10^{50} erg/s$ (e.g., Norris 2002, Ghirlanda et al. 2010).  Hence, in this MAD state, we have ample luminosity to power the GRB.  For the extremes of our parameter space (e.g. $\epsilon \sim 10^{-3}$ and $\dot{M} \sim 10^{-1}M_{\odot}/s$), this luminosity can be quite large $\sim 10^{54} erg/s$.   If the black hole is not fed angular momentum from the disk, the energy of the black hole $E_{rot} = 1/2 MR_{g}^{2}\Omega^{2}$ will be extracted in a time $E_{rot}/L_{BZ} \sim \epsilon M/\dot{M}$, where we substituted our expression for $L_{BZ}$ from equation 4 above.  This timescale can be as short as $ < 1s$ (for $\epsilon \sim 10^{-3}$ and $\dot{M} \sim 10^{-1}M_{\odot}/s$), which  is less than the lifetime of most GRBs, or as long as $\sim 100$ seconds for our more conservative values of $\epsilon$ and $\dot{M}$.  Realistically, we expect the black hole to quickly reach some equilibrium configuration in which the rotational energy extracted from the black hole is balanced by the angular momentum fed into the hole from the disk.  Tchekhovskoy et al. (2012), using GRMHD simulations of a black hole disk system, have found an equilibrium value of $a \simeq 0.07$.  However, given the unknown quantity of the equilibrium value, we adopt a mean value between Tchekhovskoy et al. 2012 and a maximally rotating hole, $a \approx 0.5$, which leads to a BZ luminosity $\sim10^{51}erg/s$ for our conservative values of $\epsilon \sim 10^{-2}$ and $\dot{M} \sim 10^{-2}M_{\odot}/s$.
 
Although black hole mass and accretion rate play a role in the estimates of the magnetic field and BZ luminosity above, our two biggest uncertainties are the parameter $\epsilon$, and the extent of the region of arrested accretion $R_{m}$.  Above, we gave arguments for why $\epsilon \sim 10^{-2}$ is a reasonable choice, but what about $R_{m}/R_{g}$?  By definition of being in a MAD state we have $R_{m}/R_{g} \geq 1$, but we do not have a good way of calculating $R_{m}$ from first principles. One way to estimate $R_{m}$ (see, e.g., Narayan et al 2003) is to get an expression for flux by integrating equation 2 out to $r = R_{m}$ ($\phi = 2 \pi \int^{R_{m}} B_{z} r dr$), and then inverting the equation to solve for $R_{m}$. We find $R_{m}/R_{g} \simeq 5 
(\phi_{29})^{4/3} (M_{5})^{-4/3}(\dot{M}_{-2})^{-2/3}(\epsilon_{10^{-2}})^{2/3}$, where $M$, $\dot{M}$, and $\epsilon$ are normalized to the values used above and $\phi$ is normalized to $10^{29} G cm^{2}$.

However, we then need an independent estimate of the magnetic flux held to the black hole to estimate $R_{m}$.  As a reasonable starting point for such a system, we can estimate the flux that arises from a magnetar field $\sim 10^{15}G$ (e.g. Kouveliotou et al. 1998), $\phi \sim 10^{29} G cm^{2}$, so that $R_{m}/R_{g} \simeq 5$.  We note that several studies (most recently M{\"o}sta et al. 2015, but see also Akiyama et al. 2003, Price \& Rosswog 2006, Zrake \& MacFadyen 2013) have shown that a dynamo can operate on very short (ms) timescales in core-collapse systems, and produce fields greater than $10^{15} G$. Alternatively, one can get an estimate of the flux from the the Blandford-Znajek luminosity expression in equation 3 above.  For an observed luminosity of $10^{51} erg s^{-1}$, and assuming an efficiency of $10\%$, we find $\phi \sim 10^{30} G cm^{2}$ which gives $R_{m}/R_{g} \sim 100$ (see also Tchekhovskoy et al. 2014, who found $\phi \sim 10^{30} G cm^{2}$ through a similar estimate). A lower efficiency will only increase the value of our magnetic flux and therefore $R_{m}$ (for example, for an efficiency of $0.5\%$, $R_{m} \sim 500 R_{g}$). The estimates here do not include the effects of rotation which may push $R_{m}$ out even further (simply from additional centrifugal support in the disk).  Hence, it is important to point out these are rough lower bounds on $R_{m}$ in a MAD state.  In what follows, we normalize $R_{m}$ to a value of $30 R_{g}$.

\subsection{Timescales}
A relevant timescale which can be compared to the observed characteristic variability time ($\sim 1s$) in the GRB is the time for gas to flow into the hole $t_{MAD} \sim R_{m}/\epsilon v_{ff}$.  Putting $R_{m}$ in terms of $R_{g}$, we find:

\medskip
\begin{equation}
t_{MAD} \simeq .3s (\frac{10^{-2}}{\epsilon})(\frac{R_{m}}{30R_{g}})^{3/2}(\frac{M}{5M_{\odot}}) 
\end{equation}
\medskip

\noindent  This timescale is near the characteristic observed variability timescale of GRBs.  It can be thought of as approximately the largest scale at which we expect variability in the disk in a MAD state, and coincides reasonably well with the 1Hz break found in the GRB power-density spectrum analysis (see \S 2 above; note that $t_{MAD}$ in equation 5 is an intrinsic time scale, and does not account for cosmological time dilation).  For each GRB, the disk will have a range of smaller scale variability less than this largest $t_{MAD}$ timescale, simply from the interchange instability spectrum (see, e.g., equation 148 of Spruit \& Taam 1990), and other instabilities in the disk.  A MAD disk itself is clumpy and subject to variable accretion (e.g. see McKinney et al. 2012), but these variability timescales are generally $<< 1s$ (for a stellar mass system) and would therefore contribute to the very high frequency variability in a GRB (and likely not be detectable in the observed light curve).  And there are of course other relevant timescales present from the emission region itself, such as radiative cooling timescales, angular timescales, etc.  All of these will come into play in the PSD of the GRB and will contribute to its behavior.   

  Although our lack of knowledge of the the microphysics details - particularly $\epsilon$ and $R_{m}$ - gives us freedom to choose parameters that coincide with the relevant observed GRB timescales, we have attempted to argue that our choices are not unreasonable for a typical GRB progenitor system, and it is encouraging that they are close to the the characteristic $1s$ timescale that comes out of the variety of GRB timing analyses (\S 2).    Figure 1 shows $t_{MAD}$ as a function of $R_{m}/R_{g}$ for a range of parameters, with $M$ between $1$ and $10 M_{\odot}$, and $\epsilon$ between $10^{-1}$ to $10^{-3}$.  The variation in progenitor parameters can accommodate the distribution of observed characteristic timescales seen in the global population of GRBs, marked by the horizontal black lines in the figure. Figure 2 shows luminosity $L_{BZ}$ as a function of time since collapse, given a realistic rotating $25 M_{\odot}$ progenitor from Heger, Langer, \& Woosley (2000), as well as the characteristic timescale $t_{MAD}$ from equation 5, for a $25 M_{\odot}$ and $75 M_{\odot}$ progenitor star. The time dependence of the luminosity (dotted line) arises directly from the variable accretion of mass and angular momentum on to the black hole $L=1/2\dot{M}(j/R_{g})^{2}$, where $j$ is the specific angular momentum (assumed to be conserved in the collapse), given the rotating progenitor model from Heger, Langer, \& Woosley 2000.  The solid and dashed lines give an indication of the bounds on $t_{MAD}$ from the masses of realistic progenitor models.

\begin{figure}
  \begin{center}
  \includegraphics[width=\columnwidth]{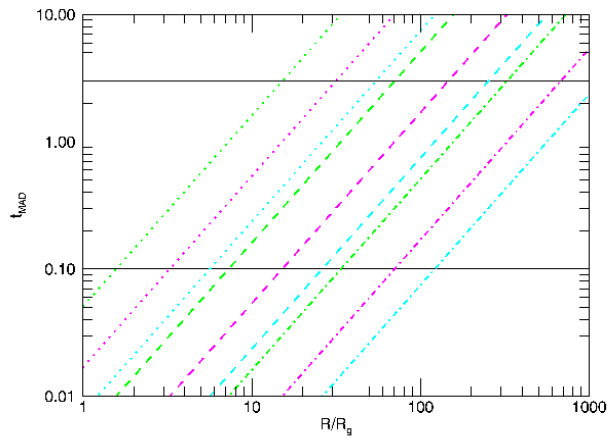}
      \caption[tvarRfinal]{Variation of the characteristic MAD timescale as a function of radius of region of arrested accretion. Dotted lines are for $\epsilon = 10^{-3}$, dashed for $\epsilon = 10^{-2}$, and dot-dashed for $\epsilon = 10^{-1}$.   Purple lines are for black hole mass $M=5M_{\odot}$, green is for $M=10M_{\odot}$, and blue is for $M=M_{\odot}$. The horizontal black lines show the approximate range of observed variability in the prompt GRB light curve}
    \label{fig:tvarR}
  \end{center}
\end{figure}

\begin{figure}
  \begin{center}
  \includegraphics[width=\columnwidth]{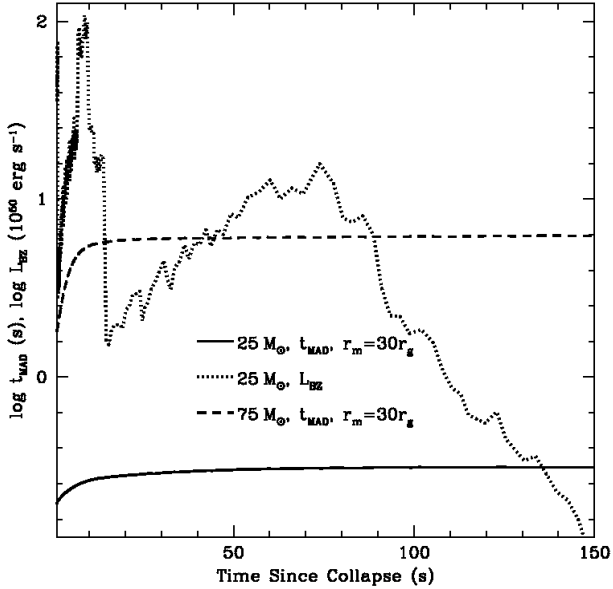}
     \caption[m25b]{Blandford-Znajek luminosity $L_{BZ}$ (dotted line) given a rotating $25M_{\odot}$ progenitor from Heger, et al. (2000), and characteristic MAD timescale $t_{MAD}$ for a $25M_{\odot}$ progenitor (solid line) and a $75M_{\odot}$ progenitor (dashed line).  The variation in $L_{BZ}$ reflects the varying angular momentum accretion rate in this model.}
    \label{fig:Lbztmad}
  \end{center}
\end{figure}

\medskip

\subsection{Ramirez-Ruiz \& Merloni Correlation}
  As mentioned in \S 2, a potentially relevant clue to the physical nature of the variability lies in the observed correlation between a quiescent time and the duration of the pulse following that quiescent time (Ramirez-Ruiz \& Merloni 2001).  In our model, this correlation could arise in two ways.   In the case in which the observed emission arises directly from modulation of the Poynting flux, the emission corresponds to the MAD state when the flux is held on to the black hole.  A quiescent time corresponds to the time in which the disk undergoes the interchange instability and the flux diffuses out into the disk.  The details of magnetic field diffusion in an accretion disk can be affected by a number of different factors (see, e.g., Guan \& Gammie 2009; Van Ballegooijen 1989; Lubow et al. 1994). The existence of a MAD state implies that the {\em timescale for diffusion of the magnetic field must be longer than the viscous timescale}, or that we have a large magnetic Prandtl number (the ratio of viscosity to resistivity).\footnote{Balbus \& Henri 2008, show that high microphysical Prantdl numbers ($> 100$) may be achieved in the inner regions of black hole-accretion disk systems. However, it may be that turbulence dominates the physics of our disk, and that is what controls the magnetic field diffusion.  In this case, we would need a high {\em turbulent} Prandtl number - the ratio turbulent viscosity to turbulent magnetic field resistivity.}  When the interchange instability occurs, the field is dissipated, and is subsequently built back up on the viscous timescale (corresponding to a quiescent time in a GRB light curve) $ \tau \propto R^{2}/\nu$, where $R$ is the radius of the disk and $\nu$ is the viscosity.  

 We can get a rough handle on this timescale using a simple Shakura-Sunyaev prescription of the disk (Shakura \& Sunyaev 1973), recognizing that this is also a model in which we parameterize our ignorance of the microphysics (here, the viscosity).  In this model,  $\nu = \alpha c_{s} H$, where $\alpha$ is a free parameter (usually between $0.01 - 0.1$), $c_{s}$ is the sound speed and $H$ is the disk height.  We can roughly relate the sound speed to the Keplerian velocity, $c_{s} \sim (H/R) \sqrt{GM/R}$.  Then we have
\begin{equation}
 \tau \simeq 1s \left(\frac{0.3}{(H/R)}\right)^{2} (\frac{0.1}{\alpha}) (\frac{R}{30R_{g}})^{3/2} (\frac{M}{5M_{\odot}}) 
\end{equation}
with a ratio of $t_{MAD} / \tau$ $\sim (\alpha/\epsilon) (H/R)^{2}$.  

Because $t_{MAD}$ and $\tau$ scale with mass and radius in the same way (assuming a constant disk scale height ratio), it is tempting to suggest that longer viscous times (which we relate to quiescent times here) imply longer MAD timescales (pulse duration episodes in our model), in accord with the observed quiescent time-pulse duration correlation found Ramirez-Ruiz \& Merloni (2001). In particular, the radial extent of the MAD disk may change from pulse to pulse, and play the primary role in producing the correlation.\footnote{Alternatively, if a shell of matter is launched during the interchange instability, it could be the launch of these (eventually) colliding shells that corresponds to the observed GRB emission.  In this case, we expect that the longer time it takes to build up a volume of mass during the arrested accretion flow (quiescent time in this case), the larger the shell of mass, and hence the longer the subsequent emission episode.}  However, this is admittedly extremely speculative and we once again emphasize that we are simply looking at general timescale dependencies, and a full treatment of the disk microphysics is necessary in order to properly get a handle on the nature of this correlation.

\section{Discussion} 
  Our model presents a picture in which the observed variability of the prompt gamma-ray burst corresponds to an accretion disk transitioning in and out of a magnetically arrested state.  We have made an estimate of a characteristic variability timescale related to free fall time in the region over which the disk is held in a magnetically arrested state, and have argued that this can be related to the observed characteristic time scale that emerges from analyses of the prompt GRB lightcurve.
One important requirement is that we assume a relatively evacuated region through which the jet propagates, so that the disk is not obscured by the surrounding medium (see, e.g.,   Thompson and Gill 2014, who argue that a hot cocoon around a jet can wash out radial variability in the jet).  Hence, a caveat for any model in which the observed variability is connected to radial variability of the outflow (originating at the central engine) is that the surrounding medium be relatively clean.

 Our assumption of a MAD state for the GRB progenitor system undoubtedly circumvents details of the microphysics of the accretion disk. There remain open questions about the specific microphysics of the disk to which concrete answers are lacking:  How can a large flux be sustained without diffusing outward?  What are the details when the interchange instability sets in?  How is the flux transported and built back up again on the black hole?  What role does turbulence play when trying to understand magnetic field advection and diffusion (e.g., Rothstein \& Lovelace 2008, Beckwith et al. 2009)?  These are all important questions for models involving magnetic field transport, and indeed conventional transport coefficients are likely inadequate to describe the physics. We refer the reader to the recent review by Lazarian et al. 2015, who discuss many of these issues. Given the many years of effort on these problems, perhaps the most promising route to getting a handle on these problems is the development of state-of-the-art numerical simulations, though current tools and resources still preclude first principles simulations.

\subsection{Late time GRB activity}  
  In our picture of a GRB, the duration of the burst is set by the lifetime of the disk.  This is an unknown parameter which depends on many things.  For various progenitors, one can look at the accretion rate as a function of time, and make an estimate of the lifetime of the disk (Murphy, Dolence \& Bromberg, in prep).  A reasonable estimate for a collapsar-type model would be a disk lifetime on the order  $\sim 100s$.  A large fraction of GRBs show features in their light curves well after the prompt gamma-ray phase at $t >> 100s$ (for example, plateaus and flares in the X-ray light curve; see, e.g., Swenson \& Roming 2014, and references therein).  Many authors have attributed this to late time activity of the central engine, based on, for example, similarities between some of the spectral and timing properties of the flares and the prompt emission (Burrows et al 2005, Fan \& Wei 2005, Zhang et al. 2006, Margutti, et al. 2011, Guidorzi et al. 2015).   Perna et al. (2006) have presented a model in which this activity is due to gravitational instabilities in a disk, causing later-time fragmented accretion episodes. Proga \& Zhang (2006) suggest a model similar to our MAD state above (although with a different set of assumptions), in which X-ray flares are caused by a disk in a magnetically arrested state. Engine-driven late time activity in a GRB poses a requirement on the central engine to remain active for hours to days after the initial prompt phase, and has been one motivation for the various magnetar models for GRBs (Usov 1992, Zhang \& Meszaros 2001, Gao \& Fan 2006, Metzger et al. 2011).  In these latter models, the initial collapse/cataclysmic event of the progenitor creates a rapidly rotating, high magnetic field, supra-massive neutron star which powers the prompt GRB.  As the neutron star spins down, it collapses to a black hole and this produces the delayed emission (plateaus and flares).

  However, we note that a long lived central engine is not required to explain the late time activity seen in these bursts.  Duffell \& MacFadyen (2015) show that plateaus arise naturally from a collapsar scenario.  The jet consists of a highly relativistic core with a baryon loaded outer shell.  The plateaus arise during the coasting phase of the external jet when the inner relativistic jet has combined with the outer jet but has not yet reached the deceleration phase.  Beniamini \& Kumar (2015) account for later time X-ray flares in a model in which mildly relativistic ejecta is emitted essentially simultaneously with the highly relativistic ejecta that produces the prompt emission.  The slower moving ejecta flares at a later time (compared to the prompt emission), when it reaches its photosphere. Van Eerten (2014) have explained observed plateaus with thick shell models without invoking extended activity of the central engine, as have several other authors employing a variety of models (e.g. Ruffini et al. 2014, Kazanas et al. 2015).

  Similarly, X-ray flares may also be a consequence of the environment of the GRB and not extended activity of the central engine.  Mesler et al. (2012, 2014) showed that massive star progenitors can produce extremely dense shells of matter that can create a flare in the GRB light curve from hours to days after the burst, depending on the location of the shell (we note that other authors have claimed density changes in the environment will {\em not} create a flaring event, e.g. Nakar \& Granot 2007, Gat et al. 2013; however, these authors either considered density changes by a factor much less than the physical models presented by Mesler et al., and/or made different assumptions about the extent of the emission region).   In addition, Margutti et al. (2011) pointed out that the main parameter driving the duration of the flare is the elapsed time since the explosion (or prompt emission episode; see their table A1).  This fits naturally into a scenario in which the flare arises from a collision with a dense circumburst shell - the further out the collision, the longer elapsed time until the flare, but also - on average - the larger the emission region and therefore duration of the flare. We note, however, that although the GRB environment can account for late time activity in the burst, these models must also explain the similarities between the timing and spectral properties of the prompt emission and the later-time flares.  This is beyond the scope of this paper and the subject of a future study.
  
 \medskip

\section{Conclusions}
  We have presented a simple model of the variability of the prompt phase of a gamma-ray burst, in which the central engine switches on and off between a magnetically arrested accretion phase.   The Poynting flux of the jet is modulated by the variable gas pressure - in other words, the ability of the accretion to anchor the flux to the event horizon, and it is this variation in Poynting flux (due to alternating in and out of a MAD state) that is directly connected to the variation in the GRB time profile.  Our simple analytic estimates show that the energetics and timescales in this model can accommodate GRB observations - significantly, the characteristic $1s$ time scale that appears in many analyses of GRB timing properties.  

  We once more point out the degeneracy of the parameters in this model when explaining GRB observations (in that we can employ a range of values to explain any given observed GRB behavior), as well as the fact that details of the microphysics of the accretion disk has been lumped into a single parameter $\epsilon$, the measure of the degree of ``arrested-ness`` of the accretion flow.  However, for reasonable values of a GRB progenitor black hole-disk system, this model naturally produces the characteristic $1s$ time scale in prompt GRB light curves.   In the future, we plan GRMHD simulations to get a better handle on the detailed physics of a MAD disk, and better test any physical connection with GRB variability.   

\medskip
 
\section*{Acknowledgements}
 We gratefully acknowledge the anonymous referee for helpful comments that improved this manuscript.  We thank Raffaella Margutti for useful discussions and for providing her thesis, and additional references. We also thank Ben Ryan and Enrico Ramirez-Ruiz for helpful comments and discussions.  This work is supported in part by the M. Hildred Blewett Fellowship of the American Physical Society, www.aps.org.  Work at LANL was done under the auspices of the National Nuclear Security Administration of the U.S. Department of Energy at Los Alamos National Laboratory, LA-UR-15-29635.


\bsp    
\label{lastpage}
\end{document}